\documentstyle[twocolumn,prl,aps]{revtex}

\begin{document}

\title{New positivity bounds on polarized
parton distributions in multicolored QCD}
\author{P.V. Pobylitsa$^{1,2}$, M.V. Polyakov$^{1,2}$}
\address{$^1$ Petersburg Nuclear Physics Institute, 188350 Gatchina,
Russia}
\address{$^2$ Institut f\"ur Theoretische Physik II,
Ruhr-Universit\"at Bochum, D-44780 Bochum, Germany}

\maketitle

\begin{abstract}
We derive new positivity bounds on spin-dependent parton distributions in
multicolored QCD. They are stronger than Soffer inequality.
We check that the new inequalities are stable under one-loop DGLAP
evolution to higher normalization points.
\end{abstract}
\vspace{0.2cm}

{\bf 1.} In spite of the vast phenomenological data providing
information about the quark and gluon distributions in nucleons we are
still far from the complete knowledge of all twist-two distributions,
especially of the spin-dependent ones.
Therefore various positivity bounds on parton distributions are very
helpful. In particular, an important role is played by Soffer inequality
\cite{Soffer-95} that constraints the transversity quark distribution which
is not measured yet.
At first sight the set of already known positivity bounds for twist-two
distributions is complete and no enhancement can be derived on general
grounds without solving the dynamics of QCD. However, if one takes QCD in
the limit of large number of colors $N_{c}$ then it is possible to obtain
new stronger inequalities.

It is well known that in the large $N_{c}$ limit \cite{Hooft-74} the baryons
are described by mean field (Hartree) equations \cite{Witten-79}, which can
be written in terms of bilocal meson fields with all possible spin and
isospin quantum numbers.
Although the corresponding effective bilocal meson action is not known, a
number of results have been obtained for the large $N_{c}$ baryons without
specifying the dynamics. Usually these results use the spin-flavor symmetry
\cite{Witten-83,Balachandran-83} of the solution of the Hartree equation
describing the nucleon and $\Delta $ resonance: this solution is invariant
under simultaneous space, spin and isospin rotations (but not invariant
under a pure flavor rotation).
Owing to this spin-flavor symmetry the baryon states (described by this mean
field solution) appear as ``rotational excitations'' with spin $S$ and
isospin $T$ connected by constraint $S=T$ \cite{Witten-83,Balachandran-83}.
In particular, this approach allows to describe nucleon ($S=T=1/2$) and
$\Delta $ resonance ($S=T=3/2$).

This picture of the large $N_{c}$ baryons is extensively used in the Skyrme
model \cite{Skyrme-61,Adkins-83} and various chiral quark-soliton models
\cite{Models}. But some general model-independent results have been
extracted for the large $N_{c}$ baryons using only the spin-flavor symmetry
\cite{Dashen-94}. This includes, for example, the identities for the mass
splitting of baryons and the large $N_{c}$ classification of spin and
isospin structures of various form factors.
In this paper we study quark distribution functions relying only on the
large $N_{c}$ limit and on the spin-flavor symmetry of the large
$N_{c}$ mean field solution.

We first present our main result --- the new positivity bounds which can be
considered as a serious enhancement of the Soffer inequality and
then give the derivation and check the stability of the bounds
under one-loop the DGLAP evolution.

\vspace{0.1cm}

{\bf 2.}
Let us first formulate the new positivity bound.
At finite number of QCD colors the unpolarized ($q_{f}$),
longitudinal polarized ($\Delta _{L}q_{f}$) and transverse polarized
($\Delta _{T}q_{f}$) quark (antiquark)
distributions are known to be constrained by
the following set of inequalities
\begin{equation} \begin{array}{c}
|\Delta _{L}q_{f}|\leq q_{f} \, ,\\
|\Delta _{T}q_{f}|\leq q_{f} \, ,\\
q_{f}+\Delta _{L}q_{f}\geq 2|\Delta _{T}q_{f}|\, .
\end{array}
\quad \quad {\rm (finite\ }N_{c}{\rm )}  \label{finite-N-c-ineq}
\end{equation}
The last inequality is known as Soffer inequality \cite{Soffer-95}. These
inequalities hold for each quark flavor $f$. In the leading order of the
large $N_{c}$ limit we have for the $u$ and $d$ distribution functions of
the proton \cite{DPPPW-96,PP-96}:
\begin{equation}
q_{u}=q_{d}\,,\quad \Delta _{L}q_{u}=-\Delta _{L}q_{d}\,,\quad \Delta
_{T}q_{u}=-\Delta _{T}q_{d}\, .
\label{q-u-q-d}
\end{equation}
Therefore the set of usual inequalities (\ref{finite-N-c-ineq}) takes the
following form in the large $N_{c}$ limit
\begin{equation}
\begin{array}{c}
|\Delta _{L}q_{u}|\leq q_{u} \, ,\\
|\Delta _{T}q_{u}|\leq q_{u} \, ,\\
q_{u}-|\Delta _{L}q_{u}|\geq 2|\Delta _{T}q_{u}|\, .
\end{array}
\quad \quad {\rm (large\, }N_{c}{\rm \, old\, bounds)}  \label{ineq-old}
\end{equation}
Note that in the last inequality (\ref{ineq-old}) we have the absolute
value of $\Delta _{L}q_{u}$ in contrast to Soffer inequality at finite $N_c$
(\ref{finite-N-c-ineq}). Actually we combined two Soffer inequalities
for $f=u,d$ (\ref{finite-N-c-ineq}) and we made use of the large
$N_c$ relation $\Delta _{L}q_{u}=-\Delta _{L}q_{d}$
(see eq.~(\ref{q-u-q-d})).

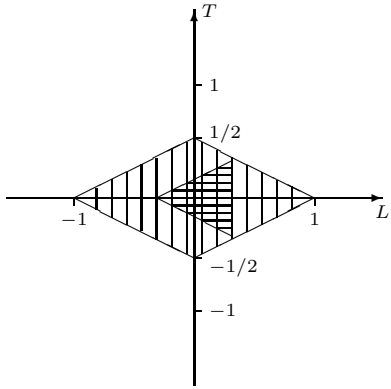
\begin{figure}[tbp]
\begin{center}
\unitlength=1mm
\linethickness{0.4pt}
\begin{picture}(55.00,55.00)
\put(30.00,5.00){\vector(0,1){50.00}}
\put(5.00,30.00){\vector(1,0){50.00}}
\put(14.00,30.00){\line(2,1){16.00}}
\put(25.00,30.00){\line(2,1){10.00}}
\put(35.00,25.00){\line(-2,1){10.00}}
\put(27.00,31.00){\line(1,0){8.00}}
\put(35.00,32.00){\line(-1,0){6.00}}
\put(31.00,33.00){\line(1,0){4.00}}
\put(35.00,34.00){\line(-1,0){2.00}}
\put(27.00,29.00){\line(1,0){8.00}}
\put(35.00,28.00){\line(-1,0){6.00}}
\put(31.00,27.00){\line(1,0){4.00}}
\put(35.00,26.00){\line(-1,0){2.00}}
\put(31.00,55.00){\makebox(0,0)[lc]{$\scriptstyle T$}}
\put(55.00,29.00){\makebox(0,0)[ct]{$\scriptstyle L$}}
\put(46.00,30.00){\line(0,-1){1.00}}
\put(46.00,28.00){\makebox(0,0)[ct]{$\scriptstyle 1$}}
\put(30.00,38.00){\line(1,0){1.00}}
\put(30.00,45.00){\line(1,0){1.00}}
\put(32.00,45.00){\makebox(0,0)[lc]{$\scriptstyle 1$}}
\put(32.00,38.00){\makebox(0,0)[lb]{$\scriptstyle 1/2$}}
\put(30.00,15.00){\line(1,0){1.00}}
\put(32.00,22.00){\makebox(0,0)[lt]{$\scriptstyle -1/2$}}
\put(32.00,15.00){\makebox(0,0)[lc]{$\scriptstyle -1$}}
\put(14.00,30.00){\line(2,-1){16.00}}
\put(30.00,38.00){\line(2,-1){16.00}}
\put(30.00,22.00){\line(2,1){16.00}}
\put(14.00,30.00){\line(0,-1){1.00}}
\put(14.00,28.00){\makebox(0,0)[ct]{$\scriptstyle -1$}}
\put(29.00,37.33){\line(0,-1){14.67}}
\put(27.00,23.50){\line(0,1){13.00}}
\put(25.00,35.33){\line(0,-1){10.83}}
\put(23.00,25.67){\line(0,1){8.83}}
\put(21.00,33.33){\line(0,-1){6.83}}
\put(19.00,27.50){\line(0,1){5.00}}
\put(17.00,31.33){\line(0,-1){2.83}}
\put(35.00,35.33){\line(0,-1){10.67}}
\put(33.00,36.33){\line(0,-1){12.83}}
\put(31.00,22.67){\line(0,1){14.67}}
\put(37.00,34.33){\line(0,-1){8.83}}
\put(39.00,26.50){\line(0,1){6.83}}
\put(41.00,32.33){\line(0,-1){4.83}}
\put(43.00,28.50){\line(0,1){3.00}}
\put(30.00,22.00){\line(1,0){1.00}}
\end{picture}
\end{center}
\caption{Soffer inequality at large $N_c$
(the rhombus hatched with vertical lines) versus new
large $N_{c}$ inequalities (the triangle hatched horizontally) in
terms of variables $L$ and $T$ (see eq.~(\ref{LT})).}
\label{Orig-fig-1}
\end{figure}

In this paper we shall derive the following
enhancement of inequalities (\ref{ineq-old})
\begin{equation}
\begin{array}{l}
\frac{1}{3}q_{u}+\Delta _{L}q_{u}\geq 2|\Delta _{T}q_{u}| \, ,\\
\   \\
\frac{1}{3}q_{u}-\Delta _{L}q_{u}\geq 0\, .
\end{array}
\quad \quad {\rm (large\, }N_{c}{\rm \, new\, bounds)}  \label{ineq-new}
\end{equation}
Let us introduce compact notations
\begin{eqnarray}
\nonumber
L&=&\frac{\Delta _{L}q_{u}}{q_{u}}=
-\frac{\Delta _{L}q_{d}}{q_{d}}+ O\left(\frac{1}{N_c} \right)
\,,\\
T&=&\frac{\Delta _{T}q_{u}}{q_{u}}
=-\frac{\Delta _{T}q_{d}}{q_{d}}+ O\left(\frac{1}{N_c} \right)\, .
\label{LT}
\end{eqnarray}
Then our new result (\ref{ineq-new}) becomes
\begin{equation}
\begin{array}{l}
\frac{1}{3}+L\geq 2|T|\, , \\
\ \\
\frac{1}{3}-L\geq 0 \, ,
\end{array}
\quad {\rm (new)}
\end{equation}
whereas the old inequalities (\ref{ineq-old}) in the limit
$N_{c}\rightarrow \infty $ result in
\begin{equation}
\begin{array}{c}
|L|\leq 1 \, ,\\
|T|\leq 1 \, ,\\
|L|+2|T|\leq 1\, .
\end{array}
\quad {\rm (old)}
\end{equation}
Graphically the old and new regions of the allowed values in the $L,T$ plane
is shown in Fig. \ref{Orig-fig-1}.

{\bf 3.} Now we turn to the derivation of our main result (\ref{ineq-new}).
In order to consider
the parton distributions in the large $N_{c}$ limit
we start from the quark correlator $\langle B_{2}|\psi (x_{2})\bar{\psi}
(x_{1})|B_{1}\rangle $. The Hartree picture of the large $N_{c}$ nucleon
described in \cite{Witten-79} leads to the following form of this
correlation function
\begin{eqnarray}
\nonumber
&&\langle B_{1},{\bf p}_{1}|\psi _{s_{1}f_{1}}^{+}(x_{1}^{0},
{\bf x}_{1})\psi _{s_{2}f_{2}}(x_{2}^{0},{\bf x}_{2})|B_{2},
{\bf p}_{2}\rangle =\\
\label{psi-psi-bar-Hartree}
&& N_{c}\int d^{3}
{\bf X}e^{i({\bf p}_{2}-{\bf p}_{1})\cdot {\bf X}}
\int dRD_{B_{1}}^{\ast }(R)D_{B_{2}}(R)\\
\nonumber
&&\times
R_{f_{2}f_{2}^{\prime
}}F_{s_{2}f_{2}^{\prime },s_{1}f_{1}^{\prime }}
\biggl(x_{1}^{0}-x_{2}^{0},
{\bf x}_{1}-{\bf X},{\bf x}_{2}-{\bf X}\biggr)
(R^{-1})_{
f_{1}^{\prime }f_{1}}
\, .
\end{eqnarray}
Here $F_{s_{2}f_{2},s_{1}f_{1}}(x_{1}^{0}-x_{2}^{0},{\bf x}_{1},{\bf x}
_{2})$ is the quark correlation function $\langle \psi^{+}(x_{1})\psi
(x_{2})\rangle $ in the background field
of the solution to the Hartree
equation, $s_{i}$ are Dirac spinor indices and $f$ are $SU(2)$ isospin
indices. The $x^{0}-y^{0}$ time dependence expresses the fact that we deal
with a static solution. Solution $
F_{s_{2}f_{2},s_{1}f_{1}}(x_{1}^{0}-x_{2}^{0},{\bf x}_{1},{\bf x}_{2})$
also violates the $SU(2)$ flavor and the 3-space translation invariance.
Therefore in eq. (\ref{psi-psi-bar-Hartree}) we used plane waves $e^{i
{\bf p}\cdot {\bf X}}$ and the rotator wave functions
$D_{B}(R)$ depending on $SU(2)$ matrix $R$
in order to construct baryon states with given spin, isospin and momentum.

Since the exact form of the effective Hartree hamiltonian corresponding to
the large $N_{c}$ limit is not known we do not know the function
$
F_{s_{2}f_{2},s_{1}f_{1}}(x_{1}^{0}-x_{2}^{0},{\bf x}_{1},{\bf x}_{2})$
either. However, following \cite{Witten-79} we assume that this solution has
the spin-flavor symmetry:
\begin{equation}
SU(2)_{\rm spin-flavor}\subset SU(2)_{\rm flavor}\otimes
SO(3,1)_{\rm Lorentz} \, .
\label{spin-flavor-sym}
\end{equation}
The transformations of $SU(2)_{\rm spin-flavor}$ perform a simultaneous
rotation both in the isospin and in the usual 3D space.
The invariance of the mean field solution under $SU(2)_{\rm spin-flavor}$
means that
\begin{eqnarray}
\nonumber
&&S_{s_{2}s_{2}^{\prime }}(R)R_{f_{2}f_{2}^{\prime }}
(R^{-1})_{f_{1}^{\prime
}f_{1}}S_{s_{1}^{\prime }s_{1}}(R^{-1})\\
&\times&
F_{s_{2}^{\prime}f_{2}^{\prime },s_{1}^{\prime }f_{1}^{\prime }}
\biggl(x^{0}_1-x^{0}_2,O(R)
{\bf x}_1,O(R){\bf x}_2\biggr)
\label{spin-flavor}
\\
\nonumber
&=&F_{s_{2}f_{2},s_{1}f_{1}}(x^{0}_1-x^{0}_2,{\bf x}_1,{\bf x}_2)\, .
\end{eqnarray}
Here we use notation $S_{s_{2}s_{2}^{\prime }}(R)$ for spin rotations and
$O(R)$ for space rotations associated with isospin matrix
$R$.

The quark distribution functions can be expressed through the
correlation function (\ref{psi-psi-bar-Hartree}) in the standard way.
Using the spin-flavor symmetry (\ref{spin-flavor}) and performing
the integration over $R$ in (\ref{psi-psi-bar-Hartree}) we easily arrives
at the large $N_c$ relations (\ref{q-u-q-d}).
Moreover, using expression
(\ref{psi-psi-bar-Hartree}) one can show that any quark
parton distribution function $f_{i}(x)$ can be represented in the large
$ N_{c}$ limit as follows
\begin{equation} f_{i}(x)={\rm Sp}\left[
\Gamma _{i}\rho (x)\right]\, .  \label{f-Gamma-rho-0}
\end{equation}
Here
$\Gamma _{i}$ and $\rho (x)$ are matrices with $SU(2)$ flavor
indices and Dirac spin indices so that we deal with the trace of $8\times 8$
matrices $(4_{spin}\times 2_{isospin}=8)$.
Matrix $\rho (x)$ can be expressed
through function
$F_{s_{2}f_{2},s_{1}f_{1}}$ (\ref{psi-psi-bar-Hartree})
as follows:
\[
\rho (x)=\frac{N_{c}}{2 \pi }\int dx^{0}\int d^{3}{\bf x}_{1}\int d^{3}
{\bf x}_{2} \int\frac{ d^3 {\bf p}}{(2\pi)^3}
e^{i {\bf p}({\bf x}_{1}-{\bf x}_{2})}
\]
\begin{equation}
\times e^{-ix^{0}(xM_{N}-p^{3})}\frac{1+\gamma ^{0}\gamma ^{3}}{2}F(x^{0},
{\bf x}_{1},{\bf x}_{2})\frac{1+\gamma ^{0}\gamma ^{3}}{2}\, .
\end{equation}
Matrix $\rho(x)$
is determined by the dynamics of the large $N_{c}$ QCD and is not known.
But $\rho (x)$ is universal for all twist-two parton distributions.

Matrices $\Gamma _{i}$ in eq.~(\ref{f-Gamma-rho-0})
are determined by the specific type of the
distribution function $f_{i}(x)$. Due to the large $N_{c}$
constraints (\ref{q-u-q-d}) we can restrict to three independent distribution
functions $f_{i}(x)$ listed in the table below with
the corresponding $8\times 8$
matrices $\Gamma_{i}$
\vspace{0.27cm}

\begin{center}
\begin{tabular}{|l|l|l|}
\hline
\raisebox{0mm}[1.2em][0.8em]{$i$}
   & $f_{i}$ & $\Gamma _{i}$ \\ \hline
\raisebox{0mm}[1.2em][0.8em]{$O$ }
  & $q_{u}+q_{d}$ & $1$ \\ \hline
\raisebox{0mm}[1.2em][0.8em]{$L$}
  & $\Delta _{L}q_{u}-\Delta _{L}q_{d}$ & $-\frac{1}{3}\gamma ^{5}\tau
^{3} $ \\ \hline
\raisebox{0mm}[1.2em][0.8em]{$T$}
 & $\Delta _{T}q_{u}-\Delta _{T}q_{d}$ & $\frac{1}{3}\gamma ^{5}\gamma
^{1}\tau ^{1}$ \\ \hline
\end{tabular}
\end{center}

\vspace{0.27cm}

Although matrix $\rho (x)$ can be found only if one solves the large $N_{c}$
QCD, one can establish its general properties:

1) $\rho $ is a hermitean matrix
\begin{equation}
\rho ^{+}=\rho \, ,
\end{equation}
positive in the matrix sense
\begin{equation}
\rho \geq 0 \, ,
\label{rho-positive-0}
\end{equation}

2) $\rho $ lives in the subspace of the projector $(1+\gamma ^{0}\gamma
^{3})/2$
\begin{equation}
\rho \gamma ^{0}\gamma ^{3}=\gamma ^{0}\gamma ^{3}\rho =\rho\, ,
\label{rho-0-3-rho}
\end{equation}

3) $\rho $ commutes with $i\gamma ^{1}\gamma ^{2}+\tau ^{3}$ due to the
spin-flavor symmetry (\ref{spin-flavor})
\begin{equation}
\lbrack \rho ,(i\gamma ^{1}\gamma ^{2}+\tau ^{3})]=0  \, .
\label{rho-K-3-commut}
\end{equation}

One could have an impression that our statements violate the Lorentz and
isotopic invariance. But actually everything is correct:
the mean field
solution for the large $N_{c}$ nucleon has the mixed spin-flavor
invariance (\ref{spin-flavor-sym}) but due to the choice of the
third axis for the boost of the infinite momentum frame
we are left only with the axial spin-isospin rotations.
As a result our eqs. (\ref{rho-0-3-rho}), (\ref{rho-K-3-commut})
are invariant only with respect
to the $U(1)$ axial spin-isospin
rotations. However, we stress that the final results for the
parton distributions are certainly both Lorentz and isospin invariant.

Using the properties (\ref{rho-0-3-rho}), (\ref{rho-K-3-commut}) of matrix
$\rho (x)$ and discrete P, C, T symmetries
one can obtain the following representation for it

\begin{equation}
\rho =\frac{1+\gamma ^{0}\gamma ^{3}}{2}\left[ c_{1}1+c_{2}\gamma ^{5}\tau
^{3}+c_{3}\gamma ^{5}(\gamma ^{1}\tau ^{1}+\gamma ^{2}\tau ^{2})\right]
\label{rho-c}
\end{equation}
with some coefficients $c_{i}$. Matrices $\gamma ^{0}\gamma ^{3}$, $\gamma
^{5}\tau ^{3}$, $\gamma ^{5}\gamma ^{1}\tau ^{1}$, $\gamma ^{5}\gamma
^{2}\tau ^{2}$ appearing in this equation commute with each other and have
eigenvalues $\pm 1$. Therefore they can be diagonalized simultaneously.
Restricting the consideration to the 4-dimensional subspace of the projector
$(1+\gamma ^{0}\gamma ^{3})/2$ (see (\ref{rho-0-3-rho}))
we can diagonalize $\rho $ as follows
\begin{equation}
\rho ={\rm diag}
(c_{1}-c_{2}+2c_{3},c_{1}+c_{2},c_{1}+c_{2},c_{1}-c_{2}-2c_{3})\, .
\nonumber
\end{equation}
The requirement of positivity (\ref{rho-positive-0}) leads to the following
constraints on coefficients $c_{i}$
\begin{equation}
c_{1}-c_{2}\geq 2|c_{3}|\, ,  \label{c-ineq-1}
\end{equation}
\begin{equation}
c_{1}+c_{2}\geq 0 \, . \label{c-ineq-2}
\end{equation}
Inserting the explicit representation for $\rho $ (\ref{rho-c}) into the
general expression for parton distributions (\ref{f-Gamma-rho-0}) we
immediately express the parton distributions $f_{i}(x)$ through
the coefficients $c_{i}$,
and the constraints (\ref{c-ineq-1}), (\ref{c-ineq-2}) on these coefficients
together with (\ref{q-u-q-d}) lead to bounds (\ref{ineq-new}) on
parton distributions.

\vspace{0.1cm}

{\bf 4.} Now we would like to check that
our new positivity bounds are preserved
by the one-loop evolution to higher normalization points
$\mu$ with DGLAP kernels taken in
the leading order of the large $N_{c}$ limit. The derivation follows the
ideas of papers \cite{Barone,Bourrely}.

In the leading order of the large $N_{c}$ limit the one-loop evolution of
the quark distributions is not affected by the gluon distribution and the
evolution equations for quark distributions listed in the above table
take the form
\begin{eqnarray}
\mu\frac{d}{d\mu}f_{O} &=&P_{||}\otimes f_{O}\, ,  \nonumber \\
\mu\frac{d}{d\mu}f_{L} &=&P_{||}\otimes f_{L}\, , \\
\mu\frac{d}{d\mu}f_{T} &=&P_{T}\otimes f_{T}\, ,  \nonumber
\end{eqnarray}
where the singlet unpolarized quark distribution $f_{O}$ and the
longitudinally polarized isovector distribution $f_{L}$ have the same
one-loop DGLAP kernel $P_{||}$.

We want to show that if for some numbers $\alpha ,\beta $ at a given
normalization point the following inequalities are satisfied
\begin{equation}
\alpha f_{L}\leq f_{O}\,,\quad \alpha f_{L}+\beta f_{T}\leq f_{O}
\label{alpha-inequality}
\end{equation}
then the evolution to higher normalization points preserves these
inequalities. Indeed, introducing functions
\begin{equation}
f_{\alpha }=f_{O}-\alpha f_{L}\,,\quad f_{\alpha ,\beta }=f_{O}-\alpha
f_{L}-\beta f_{T}\, ,
\end{equation}
we can rewrite the evolution equations as follows
\begin{equation}
\mu\frac{d}{d\mu}\left(
\begin{array}{c}
f_{\alpha ,\beta } \\
f_{\alpha }
\end{array}
\right) =\left(
\begin{array}{cc}
P_{T} & (P_{||}-P_{T}) \\
0 & P_{||}
\end{array}
\right) \otimes \left(
\begin{array}{c}
f_{\alpha ,\beta } \\
f_{\alpha }
\end{array}
\right)\, .
\end{equation}
Here all kernels are positive \cite{Artru-90}:
\begin{equation}
P_{T}>0,\quad P_{||}-P_{T}>0,\quad P_{||}>0\, ,
\end{equation}
so that the positivity of functions $F_{\alpha ,\beta }$ and $F_{\alpha }$
is preserved during the evolution to larger $\mu$
(note that
the subtraction terms in the evolution equations
do not violate this positivity \cite{Bourrely}).
This completes the
proof of
the evolution stability of inequalities (\ref{alpha-inequality}) for
arbitrary coefficients $\alpha $, $\beta $. In particular, we can choose
for $\alpha $, $\beta $ the
coefficients appearing in our bounds (\ref{ineq-new}). Thus the evolution
stability of inequalities (\ref{ineq-new}) is established.

\vspace{0.1cm}

{\bf 5.} Comparing our new bounds with the phenomenological
data one should keep
in mind that the allowed regions shown in Fig. \ref{Orig-fig-1} correspond
to the leading order of the large $N_{c}$ expansion and that the $1/N_{c}$
corrections can cause the phenomenological distribution functions to exceed
the ``triangle boundary'' of Fig. \ref{Orig-fig-1}.
Indeed, if we take the GRV \cite{GRV}
parametrization for $q_{u}+q_{d}$ and
GRSV \cite{GRSV} for $\Delta_{L}q_{u}-
\Delta _{L}q_{d}$ then we shall see that our inequality
(\ref{ineq-new})
\begin{equation}
\frac{|\Delta _{L}q_{u}-\Delta _{L}q_{d}|}{q_{u}+q_{d}}\leq \frac{1}{3}\quad
({\rm large\, }N_{c})
\end{equation}
is essentially modified by the $1/N_{c}$ corrections:
\begin{equation}
\max_{x}\frac{|\Delta _{L}
q_{u}-\Delta _{L}q_{d}|}{q_{u}+q_{d}}\sim 0.6\, .\quad
({\rm GRSV})
\end{equation}
Actually this deviation from our large $N_{c}$ bounds agrees with the fact
that in chiral quark soliton models the $1/N_{c}$ corrections
to $g_{A}$ are
known to be large \cite{Wakamatsu-93}.

The large $N_c$ inequalities derived in this paper
also hold for antiquark distributions.
Since our knowledge about $\Delta _{L}
\bar{q}_{u}-\Delta _{L}\bar{q}_{d}$ and $\Delta _{T}\bar{q}_{u}-
\Delta _{T}\bar{q}_{d}$ is still scare
the new bounds can be helpful in pinning down the
polarized antiquark distributions (note that model estimates
indicate that $1/N_c$ corrections to polarized antiquark distributions
are rather small \cite{Wakamatsu99}). The knowledge of polarized
antiquark distributions is important for the analysis of data on
semi-inclusive and Drell-Yan processes with polarized nucleons.

Apart from pure phenomenological applications we think that the new large
$N_{c}$ bounds can be used as a consistency test for various models of
polarized parton distributions.
\newline
\indent
We are grateful to A.V. Efremov, K.~Goeke, V.Yu.~Petrov,
A.~Shuvaev and O. Teryaev for inspiring discussions.
M.V.P. is thankful to Barcelona University for hospitality where a part of
this work has been done. This work was supported in parts by
RFBR grant 96-15-96764, DFG and BMFB.


\begin{thebibliography}{99}
\bibitem{Soffer-95}  J. Soffer, Phys. Rev. Lett. 74 (1995) 1292.

\bibitem{Hooft-74}  G. 't Hooft, Nucl. Phys. B72 (1974) 461.

\bibitem{Witten-79}  E. Witten, Nucl. Phys. B160 (1979) 57.

\bibitem{Witten-83}  E. Witten, Nucl. Phys. B223 (1983) 433.

\bibitem{Balachandran-83}  A.P. Balachandran, V.P Nair, S.G. Rajeev and A.
Stern, Phys. Rev. D27 (1983) 1153.

\bibitem{Skyrme-61}  T.H.R. Skyrme, Proc. Roy. Soc. A260 (1961) 127.

\bibitem{Adkins-83}  G.S. Adkins, C.R. Nappi and E. Witten, Nucl. Phys. B228
(1983) 552.

\bibitem{Models}  S. Kahana, G. Ripka and V. Soni, Nucl.\ Phys.\ A
415 (1984) 351;\newline
S. Kahana and G. Ripka, Nucl.\ Phys.\ A 429 (1984) 462.\newline
D. Diakonov and V. Petrov, Sov.\ Phys.\ JETP Lett.\ 43 (1986) 57;
\newline
D. Diakonov, V. Petrov and P. Pobylitsa, Nucl.\ Phys.\ B 306 (1988)
809; \newline
M. Wakamatsu and H. Yoshiki, Nucl.\ Phys.\ A 524 (1991) 561.

\bibitem{Dashen-94}  R. Dashen, E. Jenkins and A.V. Manohar, Phys. Rev. D49
(1994) 4713.

\bibitem{DPPPW-96}  D.I. Diakonov, V.Yu. Petrov, P.V. Pobylitsa, M.V.
Polyakov and C. Weiss, Nucl. Phys. B 480 (1996) 341.
\bibitem{PP-96} P.V. Pobylitsa and M.V.~Polyakov, Phys. Lett. B389 (1996) 350.

\bibitem{Barone}  V. Barone, Phys. Lett. B409 (1997) 499.
\bibitem{Bourrely}
C. Bourrely, J. Soffer and O. V. Teryaev,  Phys. Lett. B420 (1998) 375.
\bibitem{Artru-90}
X.~Artru and M.~Mekhi, Z.~Phys. C45 (1990) 669.

\bibitem{GRV}
M. Gl\"uck, E. Reya and A. Vogt, Z. Phys.\  C67 (1995) 433.

\bibitem{GRSV}
M. Gl\"uck, E. Reya, M. Stratmann and W. Vogelsang, Phys.\ Rev.\
D53 (1996) 4775.


\bibitem{Wakamatsu-93}  M. Wakamatsu and T. Watabe, Phys. Lett. B312 (1993)
184;\\
C.V. Christov et al., Phys. Lett. B325 (1994) 467.

\bibitem{Wakamatsu99}
M. Wakamatsu and T. Kubota, Phys. Rev. D60 (1999) 034020.
\end{thebibliography}
\end{document}